\documentclass[11pt,a4paper]{article}
\pdfoutput=1

\usepackage{amsmath,amssymb}
\usepackage{epsfig,graphicx}
\usepackage{subfigure}
\usepackage{graphicx}
\usepackage{rotating}
\usepackage{cancel}
\usepackage{bm}
\usepackage{xcolor}
\usepackage{comment}
\usepackage{cite}
\usepackage{psfrag}
\usepackage{hyperref}

\renewcommand\[{\left[}

\newcommand{\exclude}[1]{}

\def\beq{\begin{equation}}
\def\eeq{\end{equation}}

\begin{document}
\numberwithin{equation}{section}
\title{
\vspace{2.5cm} 
\Large{\textbf{Hidden Photon Dark Matter\\ in the Light of XENON1T and Stellar Cooling}\\
\vspace{0.5cm}}}

\author{Gonzalo Alonso-{\'A}lvarez$^{1}$, Fatih Ertas$^{2}$, Joerg Jaeckel$^{1}$, \\ Felix Kahlhoefer$^{2}$ and Lennert J. Thormaehlen$^{1}$\\[2ex]
\small{\em $^1$Institut f\"ur theoretische Physik, Universit\"at Heidelberg,} \\
\small{\em Philosophenweg 16, 69120 Heidelberg, Germany}\\[0.5ex]  
\small{\em $^2$Institute for Theoretical Particle Physics and Cosmology (TTK),}\\
\small{\em RWTH Aachen University, D-52056 Aachen, Germany} \\[0.8cm]}

\date{}
\maketitle

\begin{abstract}
\noindent
The low-energy electronic recoil spectrum in XENON1T provides an intriguing hint for potential new physics. At the same time, observations of horizontal branch stars favor the existence of a small amount of extra cooling compared to the one expected from the Standard Model particle content. In this note, we argue that a hidden photon with a mass of~$\sim 2.5$~keV and a kinetic mixing of~$\sim 10^{-15}$ allows for a good fit to both of these excesses. In this scenario, the signal detected in XENON1T is due to the absorption of hidden photon dark matter particles, whereas the anomalous cooling of horizontal branch stars arises from resonant production of hidden photons in the stellar interior. 
\end{abstract}

\newpage

\section{Introduction}
There are two environments known for their abundant production of light and very weakly coupled bosons: The early Universe~\cite{Preskill:1982cy,Abbott:1982af,Dine:1982ah,Nelson:2011sf,Arias:2012az} and stellar interiors~\cite{Dicus:1978fp,Ellis:1982ej,Raffelt:1985nk,Raffelt:1987yb,Raffelt:1996wa,Popov1,Popov2,Redondo:2008aa,An:2013yfc,Redondo:2013lna}.
If produced in sufficient amounts in the early Universe, light bosons may constitute the entirety of the observed dark matter~\cite{Preskill:1982cy,Abbott:1982af,Dine:1982ah,Nelson:2011sf,Arias:2012az}.
In turn, the production of these particles in a stellar interior leads to an additional energy loss that accelerates the cooling of the star~\cite{Dicus:1978fp,Ellis:1982ej,Raffelt:1985nk,Raffelt:1987yb,Raffelt:1996wa,Popov1,Popov2,Redondo:2008aa,An:2013yfc,Redondo:2013lna}.
Both effects make these particles amenable to experimental and observational tests.
Intriguingly, experiments and observations of both systems have shown (small) excesses:
XENON1T has recently reported a surplus of events in $\sim{\rm few}\times{\rm keV}$ electron recoils~\cite{Aprile:2020tmw,XENON1T_talk}, and observations of horizontal branch (HB) stars favor the existence of an extra cooling mechanism~\cite{Raffelt:1987yu,Ayala:2014pea,Giannotti:2015kwo}. While both of these could have other explanations such as a tritium component in XENON1T~\cite{Aprile:2020tmw} or an insufficient understanding of the stellar physics (as well as the statistical significance not being very large), it is nevertheless interesting to speculate.

In this brief note, we argue that hidden photons (also known as dark photons or paraphotons) that kinetically mix with the Standard Model photon may provide a simultaneous explanation for both of these observations: a single hidden photon with mass in the $m_{X}\sim 2\text{--}3\,\mathrm{keV}$ range and a kinetic mixing $\epsilon\sim 10^{-15}$ connects the hidden photon dark matter interpretation of XENON1T~\cite{Pospelov:2008jk,Aprile:2020tmw,XENON1T_talk} to the cooling excesses in HB stars~\cite{Giannotti:2015kwo}.

Another kind of light bosonic dark matter that has been suggested as the potential cause for the possible XENON1T excess are axion-like particles,\footnote{Prior to the XENON1T result, it was suggested that relatively strongly interacting but suitably light dark matter particles could cause a peak-like electron recoil signal~\cite{Smirnov:2020zwf}. It has also been proposed that the XENON1T signal could be caused by a mildly relativistic dark matter component~\cite{Kannike:2020agf}.} which were considered in the original experimental analysis~\cite{Aprile:2020tmw} and have been discussed in more detail in~\cite{Takahashi:2020bpq}. However, axion-like particle dark matter is less well suited for simultaneously explaining the XENON1T excess and the stellar cooling anomaly for the best fit region\footnote{For axion-like particles with a negligible coupling to photons, the electron coupling that fits the XENON1T result is outside of the $2\sigma$ region preferred by the stellar cooling anomalies. Note nevertheless that in addition to HB stars, the fit in~\cite{Giannotti:2017hny} includes additional data on white dwarfs and RGB stars, so a somewhat weaker fit to HB-only data may be acceptable. The agreement is also improved if axion-like particles constitute only a sub-sominant component of dark matter~\cite{Takahashi:2020bpq}.} provided in~\cite{Giannotti:2017hny}. The reason to favor the hidden photon explanation is that in the relevant region for the XENON1T signal, around $(2-3)\,{\rm keV}$, the production of hidden photons in HB stars is enhanced by a plasma resonance~\cite{Popov1,Popov2,Redondo:2008aa,An:2013yfc,Redondo:2013lna} (see below for a more detailed discussion). These stellar systems are therefore more sensitive to hidden photons than to axion-like particles in this mass range.

Furthermore and as is highlighted in~\cite{Takahashi:2020bpq}, axion-like particles with an electron coupling $g_{ae}=(5-7)\times 10^{-14}$, as required to explain the XENON1T signal, need to have extremely suppressed couplings to photons to accommodate constraints from X-ray searches (see~\cite{Cadamuro:2010cz,Irastorza:2018dyq} and references therein).
In contrast and as pointed out in this note, the most minimal hidden photon dark matter models easily evade these constraints while accounting for the XENON1T excess.

\section{Hidden photons}
Hidden photons~\cite{Okun:1982xi,Holdom:1985ag,Foot:1991kb} (cf., e.g.~\cite{Jaeckel:2013ija} for a review and further references) arise in a simple extension of the Standard Model (SM) by a U(1) vector boson under which no SM particle carries charge. Below the electroweak scale, such an extension is described by the Lagrangian
\begin{equation}
\label{eq:lagrangian}
    {\mathcal{L}}=-\frac{1}{4}(F^{\mu\nu})^2-\frac{1}{4}(X^{\mu\nu})^2-\frac{1}{2}\epsilon F^{\mu\nu}X_{\mu\nu}-\frac{1}{2}m^{2}_{X}(X^{\mu})^2-j^{\mu}A_{\mu}.
\end{equation}
In this equation, the photon (hidden photon) field $A^{\mu}$ ($X^{\mu}$) has field strength $F^{\mu\nu}$ ($X^{\mu\nu}$). An explicit mass term for the hidden photon has been included, which can be generated in a gauge-invariant way through a Higgs or St{\"u}ckelberg mechanism. The current $j^{\mu}$ summarizes all the interactions between the SM particles and the ordinary photon. Any interaction between the hidden photon and the SM takes place via the kinetic mixing term~\cite{Holdom:1985ag}.

Applying a suitable field redefinition $A^{\mu}\to A^{\mu}-\epsilon X^{\mu}$, the kinetic mixing term can be traded for a direct interaction of the hidden photon with the electrically charged SM particles $j^{\mu}A_{\mu}\to j^{\mu}(A^{\mu}-\epsilon X^{\mu})$. In particular, in this basis the interaction with electrons is explicit and has a strength $\epsilon e$, where $e$ is the electromagnetic charge. 
For the purposes of the detection in an experiment such as XENON1T, this interaction is similar to the one of a scalar or pseudoscalar axion-like particle with SM electrons via a Yukawa interaction. However, for the production in stars the situation is slightly more involved, as is discussed below.

For the present purposes, we are interested in hidden photons in the keV mass range with very small kinetic mixings of the order of $\epsilon \sim 10^{-15}$.
Constructing ultraviolet models of hidden photons that can accommodate these values is an interesting but nontrivial task.
One possibility would be to look at string-theoretic constructions~\cite{Dienes:1996zr,Lukas:1999nh,Abel:2003ue,Abel:2008ai,Jockers:2004yj,Blumenhagen:2005ga,Abel:2006qt,Goodsell:2009pi,Goodsell:2009xc,Goodsell:2010ie,Heckman:2010fh,Bullimore:2010aj,Cicoli:2011yh,Grimm:2011dx,Kerstan:2011dy,Camara:2011jg,Honecker:2011sm,Goodsell:2011wn,Marchesano:2014bia}, some of which feature the possibility of tiny kinetic mixing values (cf, e.g.~\cite{Dienes:1996zr,Abel:2003ue,Abel:2006qt,Abel:2008ai,Goodsell:2009pi,Goodsell:2009xc,Goodsell:2010ie,Cicoli:2011yh}), and small masses (see~\cite{Goodsell:2009pi,Goodsell:2009xc,Goodsell:2010ie,Cicoli:2011yh,Camara:2011jg} for some examples).  That said, a concrete model which features both a suitable mass and mixing is yet to be identified.
In the context of field theory, twin Higgs models can feature similarly minuscule kinetic  mixing parameters~\cite{Koren:2019iuv}.

\subsection*{Hidden Photon Dark Matter}
Massive hidden photons with a sufficiently small kinetic mixing constitute excellent dark matter candidates due to their feeble interactions with SM particles~\cite{Nelson:2011sf,Arias:2012az}.
For masses in the keV range or below, as is of interest here, thermal production would result in dark matter that is too warm to agree with observations of large-scale structures~\cite{Yeche:2017upn,Irsic:2017ixq,Baur:2017stq,Gilman:2019nap,Banik:2019smi}.
Furthermore, dark photons with a kinetic mixing $\epsilon\sim 10^{-15}$ interact too weakly with the visible sector for the observed abundance of dark matter to be produced through thermal processes~\cite{Redondo:2008ec}.

Therefore, a suitable abundance of hidden photons as dark matter has to be produced in a non-thermal way.
There exist a variety of non-thermal production mechanisms for hidden photon dark matter, some of which can successfully generate a cosmological population of $(2-3)\,\mathrm{keV}$ hidden photons.

Conceptually, one of the simplest ones is the misalignment mechanism~\cite{Preskill:1982cy,Abbott:1982af,Dine:1982ah} adapted to the case of hidden photons~\cite{Nelson:2011sf,Arias:2012az}.
In this setup, the field is assumed to be spatially homogeneous and is initially displaced from the minimum of its potential. At late times, when the Hubble expansion rate $H$ is smaller than $m_X$, the field oscillates around the minimum of the potential and the energy density contained in the oscillations dilutes with the volume, as befits dark matter. At early times, when $H$ is much larger than $m_X$, vectors and scalars behave differently. While a scalar field is essentially frozen in this limit, a minimally coupled vector dilutes with expansion.
Reproducing the observed dark matter density then usually requires unfeasibly large initial field values~\cite{Arias:2012az,AlonsoAlvarez:2019cgw}. This can be remedied by adding a direct coupling of the hidden photon to the curvature scalar of the form $\frac{1}{6}\kappa R (X^{\mu})^2$, where $\kappa$ is an $\mathcal{O}(1)$ coupling constant (see~\cite{Arias:2012az,AlonsoAlvarez:2019cgw} for more details). With such a coupling, the observed dark matter density can be generated for masses in the ${\rm keV}$-range.
For example, choosing some benchmark values for the field at the beginning\footnote{Here, beginning of inflation means the time at which the largest currently observable scales left the horizon, which for high scale inflation with $H_I\sim 10^{13}\,\mathrm{GeV}$ happened about $60$ e-folds before the end of inflation.} of inflation, the observed density $\Omega_{c}h^2=0.12$ is reproduced for
\begin{eqnarray}
    &m_{X}=2.5\,{\rm keV},\quad\,\, &|X_0|=10^{16}\,{\rm GeV},\qquad\qquad\quad\,\, \,\,\kappa\simeq 0.7 \, ,
    \\\nonumber    
    &m_{X}=2.5\,{\rm keV},\quad\,\, &|X_0|=M_{P}\sim 2\times 10^{18}\,{\rm GeV},\quad \kappa\simeq 0.6 \, .
\end{eqnarray}
It has also been suggested that misalignment production could be realized by invoking a non-standard gauge kinetic function~\cite{Nakayama:2019rhg} or a direct coupling to the inflaton~\cite{Nakai:2020cfw}, but these options probably run into severe isocurvature problems~\cite{Nakayama:2020rka}.

Hidden photons can also be directly produced from inflationary fluctuations~\cite{Graham:2015rva,Ema:2019yrd,AlonsoAlvarez:2019cgw,Ahmed:2020fhc}, quite independently of any initial condition. 
In the minimal scenario~\cite{Graham:2015rva}, the yield only depends on the hidden photon mass and the scale of inflation.\footnote{For scenarios in which reheating is significantly delayed, the dark matter yield can also depend on the reheating temperature~\cite{Ema:2019yrd}.} For our benchmark value, this means
\begin{equation}
    m_{X}=2.5\,{\rm keV},\quad\,\, H_{I}\sim 7\times 10^{11}\,{\rm GeV}\,.
\end{equation}
Including a coupling to $R$ as above, allows viable HP dark matter in a wider range of inflationary Hubble scales~\cite{AlonsoAlvarez:2019cgw}, 
\begin{equation}
    m_{X}=2.5\,{\rm keV},\quad\,\, H_{I}\sim 3\times 10^{12}\,{\rm GeV}-10^{14}\,{\rm GeV}, \quad\,\,  \kappa\sim 0.6-0.8.
\end{equation}
A differentiating signature of this production mechanism is the presence of very large inhomogeneities in the dark matter distribution at small scales~\cite{Graham:2015rva,AlonsoAlvarez:2019cgw}. This may offer the possibility to test this production hypothesis in the time-dependence of the signal detected at XENON1T.

Finally, a non-thermal dark matter population can also be produced if the hidden photon is coupled to a dark sector (pseudo)scalar.
The hidden photon can then be produced from a resonant decay involving an axion condensate~\cite{Agrawal:2018vin,Co:2018lka}, a dark Higgs~\cite{Dror:2018pdh}, the inflaton~\cite{Bastero-Gil:2018uel}, or a network of cosmic strings~\cite{Long:2019lwl}.
All these mechanisms allow for the production of hidden photon dark matter in the keV range with reasonable choices of parameters.

\bigskip

A pertinent question regarding viable bosonic dark matter models in the keV range is whether such particles are sufficiently long-lived. As pointed out in~\cite{Takahashi:2020bpq}, the cosmological stability of keV axion-like particles, which are prone to decay into a pair of photons, is a not a guaranteed fact. 
While the lifetimes are typically larger than the age of the Universe, stringent limits on the flux of X-rays from decaying axion-like dark matter exist~\cite{Cadamuro:2010cz,Irastorza:2018dyq}.
This means that a non-trivial suppression of the generic axionic coupling to photons is necessary in order to make axion-like particle dark matter compatible with the XENON1T signal~\cite{Takahashi:2020bpq}.

The situation of keV hidden photons is far less problematic. There are two possible decays to SM particles. The first one is the decay into three photons, which occurs with a rate~\cite{Redondo:2008ec}
\begin{equation}
     \Gamma_{X\to3\gamma}=\frac{17\alpha^4\epsilon^2}{11664000 \pi^3}\frac{m^9_{X}}{m^{8}_{e}} \simeq 5.2\times 10^{-30}\,\mathrm{Gyr}^{-1}\left( \frac{m_X}{2.5\,\mathrm{keV}} \right)^{9} \left( \frac{\epsilon}{10^{-15}} \right)^2 , 
\end{equation}
and is completely negligible in the mass range of interest.
As a matter of fact, the dominant hidden photon decay in the mass range below $\sim 10$~keV is the one into neutrinos~\cite{Ibe:2019gpv}.
For our benchmark value of $m_X$, the rate is a factor of $\sim 10$ larger than the one into photons,
\begin{equation}
     \Gamma_{X\to\nu\bar{\nu}}=\frac{\alpha\epsilon^2}{8\cos^4{\theta_W}}\frac{m^5_{X}}{m^{4}_{Z}} \simeq 1.0\times 10^{-28}\,\mathrm{Gyr}^{-1}\left( \frac{m_X}{2.5\,\mathrm{keV}} \right)^{5} \left( \frac{\epsilon}{10^{-15}} \right)^2 . 
\end{equation}
This rate does not conflict with any known constraints.

\section{Signal in XENON1T}
As reviewed above, hidden photons interact with electrons in the same way as photons, except that the strength of the coupling is suppressed by a factor of $\epsilon$. In an ambient dark matter background of hidden photons, these particles can be absorbed by xenon atoms just like photons, leading to an ionization signal in detectors like XENON1T~\cite{Pospelov:2008jk,Redondo:2008ec,Arisaka:2012pb,An:2014twa} (see Refs.~\cite{Fu:2017lfc,Akerib:2017uem,Aprile:2017lqx,Aprile:2019xxb} for earlier searches for this signal).
Due to the predicted very high abundance\footnote{As a side comment, we note that in this mass range the number density is nevertheless small enough such that typical occupation numbers are less than one and the DM is more particle- than wave-like.} of such low mass dark matter particles in the solar neighborhood, $n_{X}\sim 10^5/{\rm cm^3}\,(2.5\,{\rm keV}/m_{X})$, very small couplings are sufficient to produce a detectable signal.

The rate of dark photon dark matter absorption in a direct detection experiment per unit time and detector mass is given by~\cite{An:2014twa,Aprile:2020tmw}
\begin{equation}
R = \epsilon^2 \frac{\rho_\text{DM} }{m_X} \frac{\sigma_\gamma}{m_N}\,,
\end{equation}
where $\rho_\text{DM} = 0.3 \, \mathrm{GeV / cm^3}$ is the local DM density and $m_N$ is the target nucleus mass, while $\sigma_{\gamma}$ denotes the photoelectric cross section for the absorption of an ordinary photon by the target atoms. The resulting mono-energetic signal needs to be convoluted with the detector resolution $\sigma$, which varies between about 20\% at $E = 2\,\mathrm{keV}$ and 6\% at $E = 30 \, \mathrm{keV}$~\cite{XENON1T_talk}, leading to
\begin{equation}
\frac{\mathrm{d}R}{\mathrm{d}E} = \frac{R}{\sqrt{2 \pi}\sigma} e^{-(E-m_X)^2 / (2 \sigma^2)} \alpha(E) \; ,
\end{equation}
where $\alpha(E)$ denotes the signal efficiency.

\begin{figure}[!t]
\centering
\includegraphics[width=0.55\textwidth]{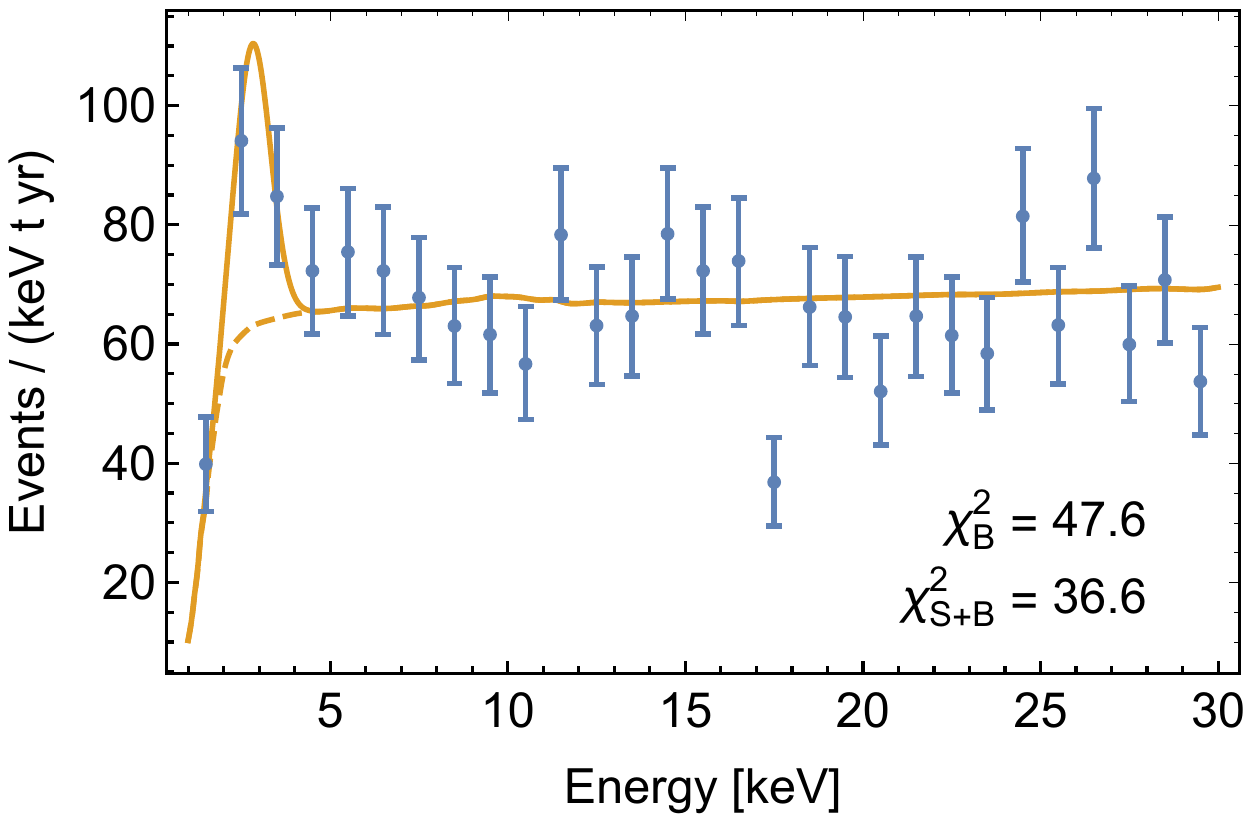}
\hspace*{1cm}
\caption{Background model from~\cite{Aprile:2020tmw} (dashed) and the best-fit signal+background (solid) for the hidden photon model compared to observations (data from~\cite{Aprile:2020tmw}).}
\label{fig:1}	
\end{figure}

We use the data from~\cite{Aprile:2020tmw}, binning the signal and background predictions in 29 equidistant bins between $1$~keV and $30$~keV in order to compare the result to data using a $\chi^2$ test statistic. For the background model $B_0$ we obtain $\chi^2_\mathrm{B} = 47.6$ (29 d.o.f.), corresponding to a $p$-value of 1.6\%. The best-fit signal hypothesis is found to be $m_X = 2.8 \, \mathrm{keV}$ and $\epsilon = 8.6 \times 10^{-16}$, giving $\chi^2_\mathrm{S+B} = 36.6$ (27 d.o.f.) and a $p$-value of 10.3\%.\footnote{When including a background contribution from a possible tritium contamination in the detector, the best-fit point shifts to $m_X = 2.7 \, \mathrm{keV}$ and $\epsilon = 6.3 \times 10^{-16}$, while the significance of the signal decreases below 2$\sigma$. We also note that there is no significant evidence for any excess at higher energies in the spectrum.} We find the local (global) significance of this signal to be of the order of $2.9\sigma$ ($\sim 2\sigma$), somewhat smaller than the value mentioned in~\cite{XENON1T_talk}, which is obtained with an unbinned profile likelihood analysis. The likely reason for this difference is that the bin width that we use is large compared to the detector resolution. We show the background model and best-fit signal prediction in Figure~\ref{fig:1}.

We also calculate the region excluded at $95\%$ C.L.\ in $m_X$-$\epsilon$ parameter space by identifying all points with $\chi^2(m_X, \epsilon) > \chi^2_\mathrm{B} + 2.70$. We furthermore define $\Delta \chi^2 = \chi^2(m_X, \epsilon) - \chi^2_\text{S+B}$ to identify the preferred parameter region around the best-fit point.\footnote{Note that we use different test statistics for the exclusion limit and the preferred parameter region and that we perform a series of one-sided test with one free parameter for the former and a two-sided test with two free parameters for the latter. Hence, the allowed parameter region at $95\%$ C.L. is different from the 95\% C.L. exclusion bound. We have checked that the 90\% C.L. exclusion bound agrees well with the one obtained by the XENON1T collaboration from an unbinned profile likelihood analysis.} At 68\% (95\%) C.L. we find $m_X \in [2.3 \, \mathrm{keV}, 3.2 \, \mathrm{keV}]$ ($m_X \in [2.0 \, \mathrm{keV}, 3.8 \, \mathrm{keV}]$).

\section{Energy loss in stars}
The core of a star, with typical temperatures of a few keV, a significant density and a large volume, constitutes an ideal source of very light bosons~\cite{Dicus:1978fp,Ellis:1982ej,Okun:1982xi,Raffelt:1985nk,Raffelt:1987yb,Raffelt:1996wa,Popov1,Popov2,Redondo:2008aa,An:2013yfc,Redondo:2013lna,Giannotti:2017hny}. 
In particular, hidden photons~\cite{Popov1,Popov2,Redondo:2008aa,An:2013yfc,Redondo:2013lna} can be produced through the couplings induced by the kinetic mixing introduced in Eq.~\eqref{eq:lagrangian}.
Once produced, these particles typically leave the star unimpeded due to their large mean free path.

One may expect the production of hidden photons via their coupling to electrons $\epsilon j^{\mu}X_{\mu}\sim \epsilon e\bar{E}\gamma^{\mu}EX_{\mu}$ (here $E$ denotes the electron field) to be very similar to the one of axion-like particles via a Yukawa interaction $\sim g_{ae}\bar{E}\gamma^{5}E$. In the stellar interior, however, plasma effects play an important role and make the physics of hidden photons comparably richer than the one of axion-like particles. In the dense medium, the photon acquires an effective mass term $\frac{1}{2}\omega^{2}_{P} A^2$ due to its interaction $j^{\mu}A_{\mu}$. Taking the hidden photon into account, the interaction shifts to $j^{\mu}(A_{\mu}-\epsilon X_{\mu})$.
This sources an additional non-diagonal term in the mass matrix for the photon--hidden photon system.
While this suppresses the production of hidden photons with a vacuum mass $m_{X}\ll \omega_{P}$, resonant conversions of transverse modes are possible when $m_{X}\sim \omega_{P}$. In this situation, the effective mixing angle can be significantly enhanced with respect to its vacuum value $\epsilon$. This leads to an enhanced production of hidden photons, making the system sensitive to even very small values of $\epsilon$, as long as $m_{X}$ lies in the vicinity of the stellar plasma mass.

A large number of very different stellar objects like the Sun \cite{Okun:1982xi,Popov1,Popov2,Redondo:2013lna,An:2014twa}, white dwarfs (WD) \cite{Isern:1992gia,Giannotti:2015kwo}, red giants (RGB) \cite{Viaux:2013lha,An:2014twa}, horizontal branch stars (HB) \cite{Raffelt:1987yu,Ayala:2014pea,An:2014twa,Giannotti:2015kwo}, blue and red giants \cite{Skillman:2002aa,McQuinn:2011bb,Friedland:2012hj} and even neutron stars \cite{Leinson:2014ioa} have been used to constrain the parameters of light bosons. In several of these objects, signs of anomalous cooling have been observed \cite{Isern:2008fs,Corsico:2012sh,Ayala:2014pea,Leinson:2014ioa}, and interpreted as a possible hint for the existence of light bosonic particles. 

The observable that is most sensitive to hidden photons in the keV range is the so-called $R$-parameter, which describes the ratio of the number of HB and upper RGB stars in globular clusters. The anomalously small value of $R$ can be explained by resonant production of hidden photons from transverse modes in HB stars.
Because of the resonant nature of the process and the higher core temperatures of RG stars, the cooling of members of the RGB are hardly affected by the presence of a keV hidden photon.

A detailed study of the constraint from the $R$-parameter on axion-like particles produced by the Primakoff effect has been carried out in Ref.~\cite{Ayala:2014pea}. For our analysis, we take the likelihood given there as a function of the Helium abundance in globular clusters (denoted $Y$) and the effective axion-photon coupling $g_{a\gamma}$. Using an updated measurement of the helium abundance, $Y = 0.2515 \pm 0.0017$~\cite{Aver:2015iza}, we calculate the profile likelihood as a function of $g_{a\gamma}$ (see Ref.~\cite{Hoof:2018ieb}). We can then use the dedicated simulations of stellar evolution from Ref.~\cite{Giannotti:2015kwo} to determine the combinations of $m_X$ and $\epsilon$ that lead to the same amount of cooling as a given value of $g_{a\gamma}$ and hence construct an $R$-parameter likelihood for hidden photons. We find that this approach leads to a $\sim 2\sigma$ preference for a non-zero cooling contribution.\footnote{Note that our approach leads to a complete degeneracy between $m_X$ and $\epsilon$. We therefore assume only one rather than two degrees of freedom when evaluating the significance and confidence levels based on HB cooling alone.}

Our results are summarised in Figure~\ref{fig:combined}, which shows the hidden photon parameter regions individually preferred by XENON1T and the $R$-parameter at $68\%$ and $95\%$ C.L. (shaded regions), as well as the respective exclusion bounds at $95\%$ C.L. and the exclusion bound from RGB stars (lines). The sudden rise of both the preferred parameter region and the exclusion limit from the $R$-parameter at $\sim$2.6~keV arises because this is the typical plasma frequency in the core of HB stars (see Ref.~\cite{An:2014twa}). For hidden photons with mass larger than this value, there is no spherical shell inside the star in which resonant production occurs and hence no significant additional cooling happens. We note that this sudden drop in the hidden photon-driven cooling will happen for slightly different values of $m_X$ for each individual star and will therefore be smeared out when considering an ensemble of HB stars representing the population in globular clusters.

Based on our implementation of the XENON1T likelihood and the $R$-parameter likelihood, we can construct a combined likelihood in order to perform a joint fit of the two anomalies. The resulting confidence regions at 68\% and 95\% C.L.\ are indicated by the red lines in Figure~\ref{fig:combined}. The red dot marks the best-fit point, which lies at $m_X = 2.55\,\mathrm{keV}$ and $\epsilon = 7.5 \times 10^{-16}$. The difference in $\chi^2$ between the best-fit point and the background-only hypothesis is $\Delta \chi^2 \approx 14$, which corresponds to a significance of $3.3\,\sigma$ (based on two degrees of freedom). When including tritium as a possible background, the global best-fit point moves to $m_X = 2.5 \, \mathrm{keV}$ and $\epsilon = 6 \times 10^{-16}$ with $\Delta \chi^2 \approx 7$, corresponding to $2.2\,\sigma$ significance.

\section{Conclusions and potential tests}

\begin{figure}[!t]
\centering
\includegraphics[width=0.55\textwidth]{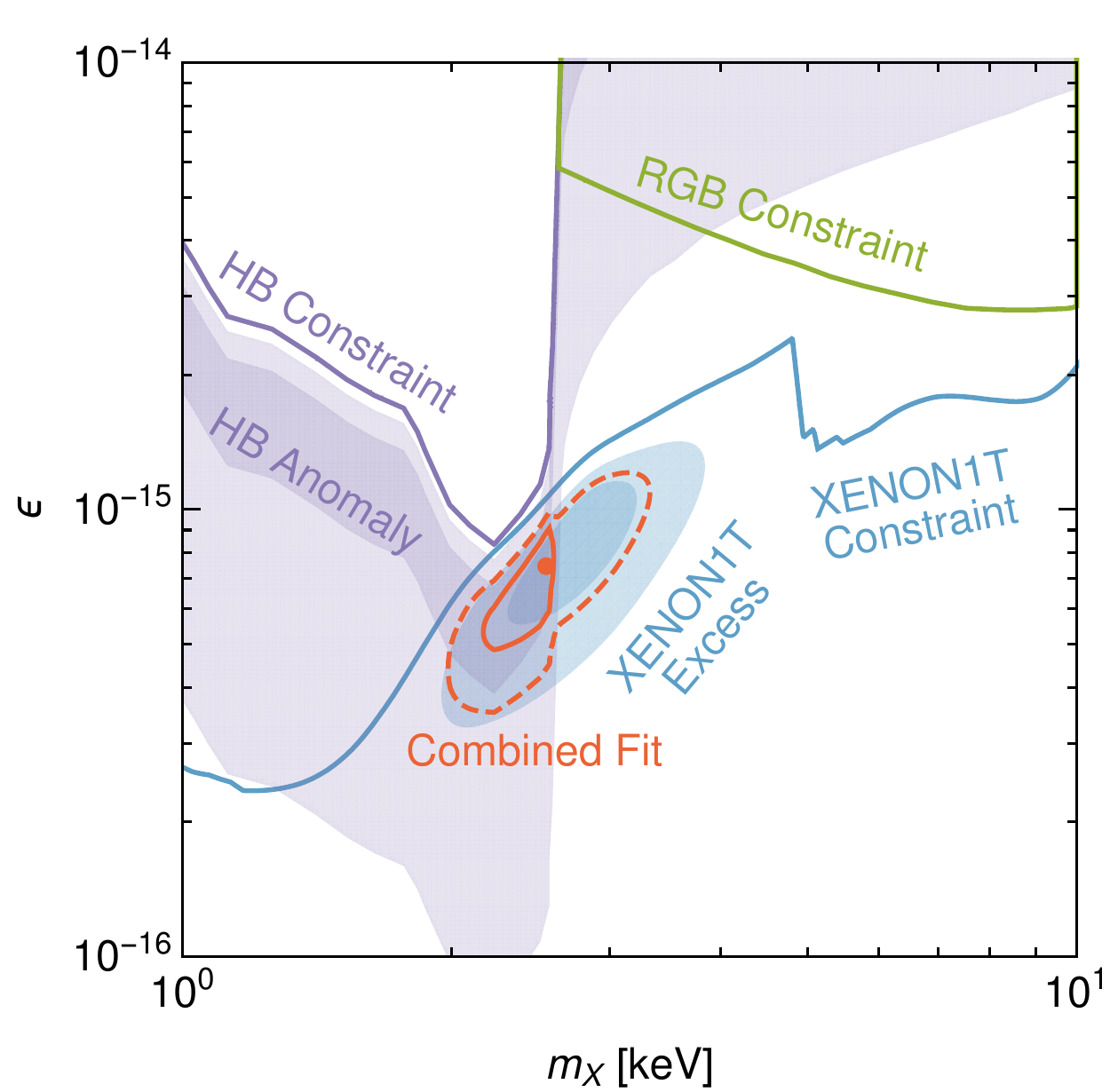}
\hspace*{1cm}
\caption{Region in the hidden photon parameter space hinted at by the XENON1T result~\cite{Aprile:2020tmw} (blue regions showing 68\% and 95\% C.L. contours) compared to the regions suggested from stellar cooling~\cite{Ayala:2014pea,Giannotti:2015kwo} (purple regions). The corresponding lines indicate 95\% C.L. exclusion limits, while the green line represents the exclusion limit from RGB stars. The red lines show the preferred parameter regions from a combined fit of the XENON1T excess and the HB anomaly at 68\% and 95\% C.L. The best-fit point is indicated by a red dot.} 
\label{fig:combined}	
\end{figure}

As can be seen from Figure~\ref{fig:combined}, the best fit region for the hidden photon dark matter fit of the XENON1T data~\cite{Aprile:2020tmw} and that of an interpretation of extra cooling in horizontal branch stars in terms of hidden photon emission~\cite{Giannotti:2015kwo} have an intriguing overlap. 
This allows for a combined explanation of both of these effects with a single hidden photon with a kinetic mixing to the Standard Model photon.

Having a simple explanation of these hints allows us also to speculate on potential complementary tests of this hypothesis. 
One possibility may be the time-dependence of the event rate in XENON1T (this was already investigated in~\cite{Aprile:2020tmw} but the results were not conclusive).
As the signal is due to the absorption of dark matter particles, the kinetic energy of them does not play a crucial role. Therefore, for a locally homogeneous dark matter distribution one expects a constant rate of events\footnote{This is in contrast to the case of a scattering of WIMPs for which a signal strongly depends on the velocity of the DM particle, leading to potential annual modulations~\cite{Drukier:1986tm}.}. This is similar to the case expected for the interpretation of axion-like particles being produced in the Sun considered by the XENON1T collaboration which is also constant up to a small annual modulation from the varying distance between Earth and Sun along the Earth's orbit~\cite{Aprile:2020tmw}. However, if hidden photons are produced from inflationary fluctuations, strong inhomogeneities are expected~\cite{Graham:2015rva,AlonsoAlvarez:2019cgw}. On small scales, this could lead to objects (similar to axion mini-clusters~\cite{Hogan:1988mp,Kolb:1993zz}) with densities about $10^4-10^5$ times higher than the average local density and a size of $\sim {\rm few}\times 100\,{\rm km}$. Such substructures would cross the detector with crossing times of the order of seconds and an encounter rate of several tens per year. 
This could perhaps be seen as a clustering of events on short time-scales, although this would probably require much larger amounts of data and probably even a larger detector such as~\cite{Aprile:2015uzo,Zhang:2018xdp,Akerib:2019fml,Aalbers:2016jon}.

On the other hand, having a relatively narrow preferred mass range for the hidden photon from the XENON1T result, also suggests doing an improved analysis of the horizontal branch hint in the resonance region. As mentioned above, this region is quite sensitive to the parameters of the stars in question. Therefore, considering the more realistic case of a distribution of horizontal branch stars and the values of their important parameters such as, e.g., the temperature, instead of taking typical values could yield extra information on the viability of the proposed interpretation. 

Given this intriguing situation, it is exciting that the next generation of dark matter experiments can conclusively shed light on the excess in XENON1T and thereby provide a test of the stellar cooling hint in horizontal branch stars.

\section*{Acknowledgements}

GA is a grateful recipient of a ``la Caixa" postgraduate fellowship from the Fundaci\'on ``la Caixa".
FE and FK are funded by the Deutsche Forschungsgemeinschaft (DFG) through the Emmy Noether Grant No.\ KA 4662/1-1. LT is funded by the Graduiertenkolleg \textit{Particle physics beyond the Standard Model} (GRK 1940).

\bibliographystyle{utphys.bst}
\bibliography{references} 

\providecommand{\href}[2]{#2}\begingroup\raggedright\begin{thebibliography}{10}

\bibitem{Preskill:1982cy}
J.~Preskill, M.~B. Wise, and F.~Wilczek, ``{Cosmology of the Invisible
  Axion},'' \href{http://dx.doi.org/10.1016/0370-2693(83)90637-8}{{\em Phys.
  Lett. B} {\bfseries 120} (1983) 127--132}.

\bibitem{Abbott:1982af}
L.~Abbott and P.~Sikivie, ``{A Cosmological Bound on the Invisible Axion},''
  \href{http://dx.doi.org/10.1016/0370-2693(83)90638-X}{{\em Phys. Lett. B}
  {\bfseries 120} (1983) 133--136}.

\bibitem{Dine:1982ah}
M.~Dine and W.~Fischler, ``{The Not So Harmless Axion},''
  \href{http://dx.doi.org/10.1016/0370-2693(83)90639-1}{{\em Phys. Lett. B}
  {\bfseries 120} (1983) 137--141}.

\bibitem{Nelson:2011sf}
A.~E. Nelson and J.~Scholtz, ``{Dark Light, Dark Matter and the Misalignment
  Mechanism},'' \href{http://dx.doi.org/10.1103/PhysRevD.84.103501}{{\em Phys.
  Rev. D} {\bfseries 84} (2011) 103501},
  \href{http://arxiv.org/abs/1105.2812}{{\ttfamily arXiv:1105.2812 [hep-ph]}}.

\bibitem{Arias:2012az}
P.~Arias, D.~Cadamuro, M.~Goodsell, J.~Jaeckel, J.~Redondo, and A.~Ringwald,
  ``{WISPy Cold Dark Matter},''
  \href{http://dx.doi.org/10.1088/1475-7516/2012/06/013}{{\em JCAP} {\bfseries
  06} (2012) 013}, \href{http://arxiv.org/abs/1201.5902}{{\ttfamily
  arXiv:1201.5902 [hep-ph]}}.

\bibitem{Dicus:1978fp}
D.~A. Dicus, E.~W. Kolb, V.~L. Teplitz, and R.~V. Wagoner, ``{Astrophysical
  Bounds on the Masses of Axions and Higgs Particles},''
  \href{http://dx.doi.org/10.1103/PhysRevD.18.1829}{{\em Phys. Rev. D}
  {\bfseries 18} (1978) 1829}.

\bibitem{Ellis:1982ej}
J.~R. Ellis and K.~A. Olive, ``{Constraints on Light Particles From Stellar
  Evolution},'' \href{http://dx.doi.org/10.1016/0550-3213(83)90104-9}{{\em
  Nucl. Phys. B} {\bfseries 223} (1983) 252--268}.

\bibitem{Raffelt:1985nk}
G.~G. Raffelt, ``{Astrophysical axion bounds diminished by screening
  effects},'' \href{http://dx.doi.org/10.1103/PhysRevD.33.897}{{\em Phys. Rev.
  D} {\bfseries 33} (1986) 897}.

\bibitem{Raffelt:1987yb}
G.~G. Raffelt and D.~S. Dearborn, ``{Bounds on Weakly Interacting Particles
  From Observational Lifetimes of Helium Burning Stars},''
  \href{http://dx.doi.org/10.1103/PhysRevD.37.549}{{\em Phys. Rev. D}
  {\bfseries 37} (1988) 549--551}.

\bibitem{Raffelt:1996wa}
G.~Raffelt, {\em {Stars as laboratories for fundamental physics}: {The
  astrophysics of neutrinos, axions, and other weakly interacting particles}}.
\newblock University of Chicago Press, 5, 1996.

\bibitem{Popov1}
V.~Popov and O.~Vasil'ev, ``{Deviations from Electrodynamics: Sun and Laser},''
  {\em EPL} {\bfseries 15} (1991) 7.

\bibitem{Popov2}
V.~Popov, ``{On the experimental search for photon mixing},'' {\em Turk. J.
  Phys.} {\bfseries 23} (1999) 943.

\bibitem{Redondo:2008aa}
J.~Redondo, ``{Helioscope Bounds on Hidden Sector Photons},''
  \href{http://dx.doi.org/10.1088/1475-7516/2008/07/008}{{\em JCAP} {\bfseries
  07} (2008) 008}, \href{http://arxiv.org/abs/0801.1527}{{\ttfamily
  arXiv:0801.1527 [hep-ph]}}.

\bibitem{An:2013yfc}
H.~An, M.~Pospelov, and J.~Pradler, ``{New stellar constraints on dark
  photons},'' \href{http://dx.doi.org/10.1016/j.physletb.2013.07.008}{{\em
  Phys. Lett. B} {\bfseries 725} (2013) 190--195},
  \href{http://arxiv.org/abs/1302.3884}{{\ttfamily arXiv:1302.3884 [hep-ph]}}.

\bibitem{Redondo:2013lna}
J.~Redondo and G.~Raffelt, ``{Solar constraints on hidden photons
  re-visited},'' \href{http://dx.doi.org/10.1088/1475-7516/2013/08/034}{{\em
  JCAP} {\bfseries 08} (2013) 034},
  \href{http://arxiv.org/abs/1305.2920}{{\ttfamily arXiv:1305.2920 [hep-ph]}}.

\bibitem{Aprile:2020tmw}
E.~Aprile {\em et~al.}, ``{Observation of Excess Electronic Recoil Events in
  XENON1T},'' \href{http://arxiv.org/abs/2006.09721}{{\ttfamily
  arXiv:2006.09721 [hep-ex]}}.

\bibitem{XENON1T_talk}
{\bfseries XENON1T} Collaboration, E.~Shockley, ``{Search for New Physics with
  Electronic-Recoil Events in XENON1T},'' 2020.
\newblock Presented as LNGS Webinar on 17 June.

\bibitem{Raffelt:1987yu}
G.~G. Raffelt and D.~S. Dearborn, ``{Bounds on Hadronic Axions From Stellar
  Evolution},'' \href{http://dx.doi.org/10.1103/PhysRevD.36.2211}{{\em Phys.
  Rev. D} {\bfseries 36} (1987) 2211}.

\bibitem{Ayala:2014pea}
A.~Ayala, I.~Domínguez, M.~Giannotti, A.~Mirizzi, and O.~Straniero,
  ``{Revisiting the bound on axion-photon coupling from Globular Clusters},''
  \href{http://dx.doi.org/10.1103/PhysRevLett.113.191302}{{\em Phys. Rev.
  Lett.} {\bfseries 113} no.~19, (2014) 191302},
  \href{http://arxiv.org/abs/1406.6053}{{\ttfamily arXiv:1406.6053
  [astro-ph.SR]}}.

\bibitem{Giannotti:2015kwo}
M.~Giannotti, I.~Irastorza, J.~Redondo, and A.~Ringwald, ``{Cool WISPs for
  stellar cooling excesses},''
  \href{http://dx.doi.org/10.1088/1475-7516/2016/05/057}{{\em JCAP} {\bfseries
  05} (2016) 057}, \href{http://arxiv.org/abs/1512.08108}{{\ttfamily
  arXiv:1512.08108 [astro-ph.HE]}}.

\bibitem{Pospelov:2008jk}
M.~Pospelov, A.~Ritz, and M.~B. Voloshin, ``{Bosonic super-WIMPs as keV-scale
  dark matter},'' \href{http://dx.doi.org/10.1103/PhysRevD.78.115012}{{\em
  Phys. Rev. D} {\bfseries 78} (2008) 115012},
  \href{http://arxiv.org/abs/0807.3279}{{\ttfamily arXiv:0807.3279 [hep-ph]}}.

\bibitem{Smirnov:2020zwf}
J.~Smirnov and J.~F. Beacom, ``{Co-SIMP Miracle},''
  \href{http://arxiv.org/abs/2002.04038}{{\ttfamily arXiv:2002.04038
  [hep-ph]}}.

\bibitem{Kannike:2020agf}
K.~Kannike, M.~Raidal, H.~Veermäe, A.~Strumia, and D.~Teresi, ``{Dark Matter
  and the XENON1T electron recoil excess},''
  \href{http://arxiv.org/abs/2006.10735}{{\ttfamily arXiv:2006.10735
  [hep-ph]}}.

\bibitem{Takahashi:2020bpq}
F.~Takahashi, M.~Yamada, and W.~Yin, ``{XENON1T anomaly from anomaly-free ALP
  dark matter},'' \href{http://arxiv.org/abs/2006.10035}{{\ttfamily
  arXiv:2006.10035 [hep-ph]}}.

\bibitem{Giannotti:2017hny}
M.~Giannotti, I.~G. Irastorza, J.~Redondo, A.~Ringwald, and K.~Saikawa,
  ``{Stellar Recipes for Axion Hunters},''
  \href{http://dx.doi.org/10.1088/1475-7516/2017/10/010}{{\em JCAP} {\bfseries
  10} (2017) 010}, \href{http://arxiv.org/abs/1708.02111}{{\ttfamily
  arXiv:1708.02111 [hep-ph]}}.

\bibitem{Cadamuro:2010cz}
D.~Cadamuro, S.~Hannestad, G.~Raffelt, and J.~Redondo, ``{Cosmological bounds
  on sub-MeV mass axions},''
  \href{http://dx.doi.org/10.1088/1475-7516/2011/02/003}{{\em JCAP} {\bfseries
  02} (2011) 003}, \href{http://arxiv.org/abs/1011.3694}{{\ttfamily
  arXiv:1011.3694 [hep-ph]}}.

\bibitem{Irastorza:2018dyq}
I.~G. Irastorza and J.~Redondo, ``{New experimental approaches in the search
  for axion-like particles},''
  \href{http://dx.doi.org/10.1016/j.ppnp.2018.05.003}{{\em Prog. Part. Nucl.
  Phys.} {\bfseries 102} (2018) 89--159},
  \href{http://arxiv.org/abs/1801.08127}{{\ttfamily arXiv:1801.08127
  [hep-ph]}}.

\bibitem{Okun:1982xi}
L.~Okun, ``{Limits on electrodynamics: paraphotons?},'' {\em Sov. Phys. JETP}
  {\bfseries 56} (1982) 502.

\bibitem{Holdom:1985ag}
B.~Holdom, ``{Two U(1)'s and Epsilon Charge Shifts},''
  \href{http://dx.doi.org/10.1016/0370-2693(86)91377-8}{{\em Phys. Lett. B}
  {\bfseries 166} (1986) 196--198}.

\bibitem{Foot:1991kb}
R.~Foot and X.-G. He, ``{Comment on Z Z-prime mixing in extended gauge
  theories},'' \href{http://dx.doi.org/10.1016/0370-2693(91)90901-2}{{\em Phys.
  Lett. B} {\bfseries 267} (1991) 509--512}.

\bibitem{Jaeckel:2013ija}
J.~Jaeckel, ``{A force beyond the Standard Model - Status of the quest for
  hidden photons},'' {\em Frascati Phys. Ser.} {\bfseries 56} (2012) 172--192,
  \href{http://arxiv.org/abs/1303.1821}{{\ttfamily arXiv:1303.1821 [hep-ph]}}.

\bibitem{Dienes:1996zr}
K.~R. Dienes, C.~F. Kolda, and J.~March-Russell, ``{Kinetic mixing and the
  supersymmetric gauge hierarchy},''
  \href{http://dx.doi.org/10.1016/S0550-3213(97)00173-9}{{\em Nucl. Phys. B}
  {\bfseries 492} (1997) 104--118},
  \href{http://arxiv.org/abs/hep-ph/9610479}{{\ttfamily arXiv:hep-ph/9610479}}.

\bibitem{Lukas:1999nh}
A.~Lukas and K.~Stelle, ``{Heterotic anomaly cancellation in
  five-dimensions},''
  \href{http://dx.doi.org/10.1088/1126-6708/2000/01/010}{{\em JHEP} {\bfseries
  01} (2000) 010}, \href{http://arxiv.org/abs/hep-th/9911156}{{\ttfamily
  arXiv:hep-th/9911156}}.

\bibitem{Abel:2003ue}
S.~Abel and B.~Schofield, ``{Brane anti-brane kinetic mixing, millicharged
  particles and SUSY breaking},''
  \href{http://dx.doi.org/10.1016/j.nuclphysb.2004.02.037}{{\em Nucl. Phys. B}
  {\bfseries 685} (2004) 150--170},
  \href{http://arxiv.org/abs/hep-th/0311051}{{\ttfamily arXiv:hep-th/0311051}}.

\bibitem{Abel:2008ai}
S.~Abel, M.~Goodsell, J.~Jaeckel, V.~Khoze, and A.~Ringwald, ``{Kinetic Mixing
  of the Photon with Hidden U(1)s in String Phenomenology},''
  \href{http://dx.doi.org/10.1088/1126-6708/2008/07/124}{{\em JHEP} {\bfseries
  07} (2008) 124}, \href{http://arxiv.org/abs/0803.1449}{{\ttfamily
  arXiv:0803.1449 [hep-ph]}}.

\bibitem{Jockers:2004yj}
H.~Jockers and J.~Louis, ``{The Effective action of D7-branes in N = 1
  Calabi-Yau orientifolds},''
  \href{http://dx.doi.org/10.1016/j.nuclphysb.2004.11.009}{{\em Nucl. Phys. B}
  {\bfseries 705} (2005) 167--211},
  \href{http://arxiv.org/abs/hep-th/0409098}{{\ttfamily arXiv:hep-th/0409098}}.

\bibitem{Blumenhagen:2005ga}
R.~Blumenhagen, G.~Honecker, and T.~Weigand, ``{Loop-corrected
  compactifications of the heterotic string with line bundles},''
  \href{http://dx.doi.org/10.1088/1126-6708/2005/06/020}{{\em JHEP} {\bfseries
  06} (2005) 020}, \href{http://arxiv.org/abs/hep-th/0504232}{{\ttfamily
  arXiv:hep-th/0504232}}.

\bibitem{Abel:2006qt}
S.~A. Abel, J.~Jaeckel, V.~V. Khoze, and A.~Ringwald, ``{Illuminating the
  Hidden Sector of String Theory by Shining Light through a Magnetic Field},''
  \href{http://dx.doi.org/10.1016/j.physletb.2008.03.076}{{\em Phys. Lett. B}
  {\bfseries 666} (2008) 66--70},
  \href{http://arxiv.org/abs/hep-ph/0608248}{{\ttfamily arXiv:hep-ph/0608248}}.

\bibitem{Goodsell:2009pi}
M.~Goodsell,
  \href{http://dx.doi.org/10.3204/DESY-PROC-2009-05/goodsell\_mark}{``{Light
  Hidden U(1)s from String Theory},''} in {\em {5th Patras Workshop on Axions,
  WIMPs and WISPs}}, pp.~165--168.
\newblock 12, 2009.
\newblock \href{http://arxiv.org/abs/0912.4206}{{\ttfamily arXiv:0912.4206
  [hep-th]}}.

\bibitem{Goodsell:2009xc}
M.~Goodsell, J.~Jaeckel, J.~Redondo, and A.~Ringwald, ``{Naturally Light Hidden
  Photons in LARGE Volume String Compactifications},''
  \href{http://dx.doi.org/10.1088/1126-6708/2009/11/027}{{\em JHEP} {\bfseries
  11} (2009) 027}, \href{http://arxiv.org/abs/0909.0515}{{\ttfamily
  arXiv:0909.0515 [hep-ph]}}.

\bibitem{Goodsell:2010ie}
M.~Goodsell and A.~Ringwald, ``{Light Hidden-Sector U(1)s in String
  Compactifications},'' \href{http://dx.doi.org/10.1002/prop.201000026}{{\em
  Fortsch. Phys.} {\bfseries 58} (2010) 716--720},
  \href{http://arxiv.org/abs/1002.1840}{{\ttfamily arXiv:1002.1840 [hep-th]}}.

\bibitem{Heckman:2010fh}
J.~J. Heckman and C.~Vafa, ``{An Exceptional Sector for F-theory GUTs},''
  \href{http://dx.doi.org/10.1103/PhysRevD.83.026006}{{\em Phys. Rev. D}
  {\bfseries 83} (2011) 026006},
  \href{http://arxiv.org/abs/1006.5459}{{\ttfamily arXiv:1006.5459 [hep-th]}}.

\bibitem{Bullimore:2010aj}
M.~Bullimore, J.~P. Conlon, and L.~T. Witkowski, ``{Kinetic mixing of U(1)s for
  local string models},'' \href{http://dx.doi.org/10.1007/JHEP11(2010)142}{{\em
  JHEP} {\bfseries 11} (2010) 142},
  \href{http://arxiv.org/abs/1009.2380}{{\ttfamily arXiv:1009.2380 [hep-th]}}.

\bibitem{Cicoli:2011yh}
M.~Cicoli, M.~Goodsell, J.~Jaeckel, and A.~Ringwald, ``{Testing String Vacua in
  the Lab: From a Hidden CMB to Dark Forces in Flux Compactifications},''
  \href{http://dx.doi.org/10.1007/JHEP07(2011)114}{{\em JHEP} {\bfseries 07}
  (2011) 114}, \href{http://arxiv.org/abs/1103.3705}{{\ttfamily arXiv:1103.3705
  [hep-th]}}.

\bibitem{Grimm:2011dx}
T.~W. Grimm and D.~Vieira~Lopes, ``{The N=1 effective actions of D-branes in
  Type IIA and IIB orientifolds},''
  \href{http://dx.doi.org/10.1016/j.nuclphysb.2011.10.019}{{\em Nucl. Phys. B}
  {\bfseries 855} (2012) 639--694},
  \href{http://arxiv.org/abs/1104.2328}{{\ttfamily arXiv:1104.2328 [hep-th]}}.

\bibitem{Kerstan:2011dy}
M.~Kerstan and T.~Weigand, ``{The Effective action of D6-branes in N=1 type IIA
  orientifolds},'' \href{http://dx.doi.org/10.1007/JHEP06(2011)105}{{\em JHEP}
  {\bfseries 06} (2011) 105}, \href{http://arxiv.org/abs/1104.2329}{{\ttfamily
  arXiv:1104.2329 [hep-th]}}.

\bibitem{Camara:2011jg}
P.~G. Camara, L.~E. Ibanez, and F.~Marchesano, ``{RR photons},''
  \href{http://dx.doi.org/10.1007/JHEP09(2011)110}{{\em JHEP} {\bfseries 09}
  (2011) 110}, \href{http://arxiv.org/abs/1106.0060}{{\ttfamily arXiv:1106.0060
  [hep-th]}}.

\bibitem{Honecker:2011sm}
G.~Honecker, ``{Kaehler metrics and gauge kinetic functions for intersecting
  D6-branes on toroidal orbifolds - The complete perturbative story},''
  \href{http://dx.doi.org/10.1002/prop.201100087}{{\em Fortsch. Phys.}
  {\bfseries 60} (2012) 243--326},
  \href{http://arxiv.org/abs/1109.3192}{{\ttfamily arXiv:1109.3192 [hep-th]}}.

\bibitem{Goodsell:2011wn}
M.~Goodsell, S.~Ramos-Sanchez, and A.~Ringwald, ``{Kinetic Mixing of U(1)s in
  Heterotic Orbifolds},'' \href{http://dx.doi.org/10.1007/JHEP01(2012)021}{{\em
  JHEP} {\bfseries 01} (2012) 021},
  \href{http://arxiv.org/abs/1110.6901}{{\ttfamily arXiv:1110.6901 [hep-th]}}.

\bibitem{Marchesano:2014bia}
F.~Marchesano, D.~Regalado, and G.~Zoccarato, ``{U(1) mixing and D-brane linear
  equivalence},'' \href{http://dx.doi.org/10.1007/JHEP08(2014)157}{{\em JHEP}
  {\bfseries 08} (2014) 157}, \href{http://arxiv.org/abs/1406.2729}{{\ttfamily
  arXiv:1406.2729 [hep-th]}}.

\bibitem{Koren:2019iuv}
S.~Koren and R.~McGehee, ``{Freezing-in twin dark matter},''
  \href{http://dx.doi.org/10.1103/PhysRevD.101.055024}{{\em Phys. Rev. D}
  {\bfseries 101} no.~5, (2020) 055024},
  \href{http://arxiv.org/abs/1908.03559}{{\ttfamily arXiv:1908.03559
  [hep-ph]}}.

\bibitem{Yeche:2017upn}
C.~Yèche, N.~Palanque-Delabrouille, J.~Baur, and H.~du~Mas~des Bourboux,
  ``{Constraints on neutrino masses from Lyman-alpha forest power spectrum with
  BOSS and XQ-100},''
  \href{http://dx.doi.org/10.1088/1475-7516/2017/06/047}{{\em JCAP} {\bfseries
  06} (2017) 047}, \href{http://arxiv.org/abs/1702.03314}{{\ttfamily
  arXiv:1702.03314 [astro-ph.CO]}}.

\bibitem{Irsic:2017ixq}
V.~Ir\v~si\v c {\em et~al.}, ``{New Constraints on the free-streaming of warm
  dark matter from intermediate and small scale Lyman-$\alpha$ forest data},''
  \href{http://dx.doi.org/10.1103/PhysRevD.96.023522}{{\em Phys. Rev. D}
  {\bfseries 96} no.~2, (2017) 023522},
  \href{http://arxiv.org/abs/1702.01764}{{\ttfamily arXiv:1702.01764
  [astro-ph.CO]}}.

\bibitem{Baur:2017stq}
J.~Baur, N.~Palanque-Delabrouille, C.~Yeche, A.~Boyarsky, O.~Ruchayskiy,
  {\'E}.~Armengaud, and J.~Lesgourgues, ``{Constraints from Ly-$\alpha$ forests
  on non-thermal dark matter including resonantly-produced sterile
  neutrinos},'' \href{http://dx.doi.org/10.1088/1475-7516/2017/12/013}{{\em
  JCAP} {\bfseries 12} (2017) 013},
  \href{http://arxiv.org/abs/1706.03118}{{\ttfamily arXiv:1706.03118
  [astro-ph.CO]}}.

\bibitem{Gilman:2019nap}
D.~Gilman, S.~Birrer, A.~Nierenberg, T.~Treu, X.~Du, and A.~Benson, ``{Warm
  dark matter chills out: constraints on the halo mass function and the
  free-streaming length of dark matter with eight quadruple-image strong
  gravitational lenses},'' \href{http://dx.doi.org/10.1093/mnras/stz3480}{{\em
  Mon. Not. Roy. Astron. Soc.} {\bfseries 491} no.~4, (2020) 6077--6101},
  \href{http://arxiv.org/abs/1908.06983}{{\ttfamily arXiv:1908.06983
  [astro-ph.CO]}}.

\bibitem{Banik:2019smi}
N.~Banik, J.~Bovy, G.~Bertone, D.~Erkal, and T.~de~Boer, ``{Novel constraints
  on the particle nature of dark matter from stellar streams},''
  \href{http://arxiv.org/abs/1911.02663}{{\ttfamily arXiv:1911.02663
  [astro-ph.GA]}}.

\bibitem{Redondo:2008ec}
J.~Redondo and M.~Postma, ``{Massive hidden photons as lukewarm dark matter},''
  \href{http://dx.doi.org/10.1088/1475-7516/2009/02/005}{{\em JCAP} {\bfseries
  02} (2009) 005}, \href{http://arxiv.org/abs/0811.0326}{{\ttfamily
  arXiv:0811.0326 [hep-ph]}}.

\bibitem{AlonsoAlvarez:2019cgw}
G.~Alonso-Álvarez, T.~Hugle, and J.~Jaeckel, ``{Misalignment \& Co.:
  (Pseudo-)scalar and vector dark matter with curvature couplings},''
  \href{http://dx.doi.org/10.1088/1475-7516/2020/02/014}{{\em JCAP} {\bfseries
  02} (2020) 014}, \href{http://arxiv.org/abs/1905.09836}{{\ttfamily
  arXiv:1905.09836 [hep-ph]}}.

\bibitem{Nakayama:2019rhg}
K.~Nakayama, ``{Vector Coherent Oscillation Dark Matter},''
  \href{http://dx.doi.org/10.1088/1475-7516/2019/10/019}{{\em JCAP} {\bfseries
  10} (2019) 019}, \href{http://arxiv.org/abs/1907.06243}{{\ttfamily
  arXiv:1907.06243 [hep-ph]}}.

\bibitem{Nakai:2020cfw}
Y.~Nakai, R.~Namba, and Z.~Wang, ``{Light Dark Photon Dark Matter from
  Inflation},'' \href{http://arxiv.org/abs/2004.10743}{{\ttfamily
  arXiv:2004.10743 [hep-ph]}}.

\bibitem{Nakayama:2020rka}
K.~Nakayama, ``{Constraint on Vector Coherent Oscillation Dark Matter with
  Kinetic Function},'' \href{http://arxiv.org/abs/2004.10036}{{\ttfamily
  arXiv:2004.10036 [hep-ph]}}.

\bibitem{Graham:2015rva}
P.~W. Graham, J.~Mardon, and S.~Rajendran, ``{Vector Dark Matter from
  Inflationary Fluctuations},''
  \href{http://dx.doi.org/10.1103/PhysRevD.93.103520}{{\em Phys. Rev. D}
  {\bfseries 93} no.~10, (2016) 103520},
  \href{http://arxiv.org/abs/1504.02102}{{\ttfamily arXiv:1504.02102
  [hep-ph]}}.

\bibitem{Ema:2019yrd}
Y.~Ema, K.~Nakayama, and Y.~Tang, ``{Production of Purely Gravitational Dark
  Matter: The Case of Fermion and Vector Boson},''
  \href{http://dx.doi.org/10.1007/JHEP07(2019)060}{{\em JHEP} {\bfseries 07}
  (2019) 060}, \href{http://arxiv.org/abs/1903.10973}{{\ttfamily
  arXiv:1903.10973 [hep-ph]}}.

\bibitem{Ahmed:2020fhc}
A.~Ahmed, B.~Grzadkowski, and A.~Socha, ``{Gravitational production of vector
  dark matter},'' \href{http://arxiv.org/abs/2005.01766}{{\ttfamily
  arXiv:2005.01766 [hep-ph]}}.

\bibitem{Agrawal:2018vin}
P.~Agrawal, N.~Kitajima, M.~Reece, T.~Sekiguchi, and F.~Takahashi, ``{Relic
  Abundance of Dark Photon Dark Matter},''
  \href{http://dx.doi.org/10.1016/j.physletb.2019.135136}{{\em Phys. Lett. B}
  {\bfseries 801} (2020) 135136},
  \href{http://arxiv.org/abs/1810.07188}{{\ttfamily arXiv:1810.07188
  [hep-ph]}}.

\bibitem{Co:2018lka}
R.~T. Co, A.~Pierce, Z.~Zhang, and Y.~Zhao, ``{Dark Photon Dark Matter Produced
  by Axion Oscillations},''
  \href{http://dx.doi.org/10.1103/PhysRevD.99.075002}{{\em Phys. Rev. D}
  {\bfseries 99} no.~7, (2019) 075002},
  \href{http://arxiv.org/abs/1810.07196}{{\ttfamily arXiv:1810.07196
  [hep-ph]}}.

\bibitem{Dror:2018pdh}
J.~A. Dror, K.~Harigaya, and V.~Narayan, ``{Parametric Resonance Production of
  Ultralight Vector Dark Matter},''
  \href{http://dx.doi.org/10.1103/PhysRevD.99.035036}{{\em Phys. Rev. D}
  {\bfseries 99} no.~3, (2019) 035036},
  \href{http://arxiv.org/abs/1810.07195}{{\ttfamily arXiv:1810.07195
  [hep-ph]}}.

\bibitem{Bastero-Gil:2018uel}
M.~Bastero-Gil, J.~Santiago, L.~Ubaldi, and R.~Vega-Morales, ``{Vector dark
  matter production at the end of inflation},''
  \href{http://dx.doi.org/10.1088/1475-7516/2019/04/015}{{\em JCAP} {\bfseries
  04} (2019) 015}, \href{http://arxiv.org/abs/1810.07208}{{\ttfamily
  arXiv:1810.07208 [hep-ph]}}.

\bibitem{Long:2019lwl}
A.~J. Long and L.-T. Wang, ``{Dark Photon Dark Matter from a Network of Cosmic
  Strings},'' \href{http://dx.doi.org/10.1103/PhysRevD.99.063529}{{\em Phys.
  Rev. D} {\bfseries 99} no.~6, (2019) 063529},
  \href{http://arxiv.org/abs/1901.03312}{{\ttfamily arXiv:1901.03312
  [hep-ph]}}.

\bibitem{Ibe:2019gpv}
M.~Ibe, S.~Kobayashi, Y.~Nakayama, and S.~Shirai, ``{Cosmological constraint on
  dark photon from N$_{eff}$},''
  \href{http://dx.doi.org/10.1007/JHEP04(2020)009}{{\em JHEP} {\bfseries 04}
  (2020) 009}, \href{http://arxiv.org/abs/1912.12152}{{\ttfamily
  arXiv:1912.12152 [hep-ph]}}.

\bibitem{Arisaka:2012pb}
K.~Arisaka, P.~Beltrame, C.~Ghag, J.~Kaidi, K.~Lung, A.~Lyashenko, R.~Peccei,
  P.~Smith, and K.~Ye, ``{Expected Sensitivity to Galactic/Solar Axions and
  Bosonic Super-WIMPs based on the Axio-electric Effect in Liquid Xenon Dark
  Matter Detectors},''
  \href{http://dx.doi.org/10.1016/j.astropartphys.2012.12.009}{{\em Astropart.
  Phys.} {\bfseries 44} (2013) 59--67},
  \href{http://arxiv.org/abs/1209.3810}{{\ttfamily arXiv:1209.3810
  [astro-ph.CO]}}.

\bibitem{An:2014twa}
H.~An, M.~Pospelov, J.~Pradler, and A.~Ritz, ``{Direct Detection Constraints on
  Dark Photon Dark Matter},''
  \href{http://dx.doi.org/10.1016/j.physletb.2015.06.018}{{\em Phys. Lett. B}
  {\bfseries 747} (2015) 331--338},
  \href{http://arxiv.org/abs/1412.8378}{{\ttfamily arXiv:1412.8378 [hep-ph]}}.

\bibitem{Fu:2017lfc}
{\bfseries PandaX} Collaboration, C.~Fu {\em et~al.}, ``{Limits on Axion
  Couplings from the First 80 Days of Data of the PandaX-II Experiment},''
  \href{http://dx.doi.org/10.1103/PhysRevLett.119.181806}{{\em Phys. Rev.
  Lett.} {\bfseries 119} no.~18, (2017) 181806},
  \href{http://arxiv.org/abs/1707.07921}{{\ttfamily arXiv:1707.07921
  [hep-ex]}}.

\bibitem{Akerib:2017uem}
{\bfseries LUX} Collaboration, D.~Akerib {\em et~al.}, ``{First Searches for
  Axions and Axionlike Particles with the LUX Experiment},''
  \href{http://dx.doi.org/10.1103/PhysRevLett.118.261301}{{\em Phys. Rev.
  Lett.} {\bfseries 118} no.~26, (2017) 261301},
  \href{http://arxiv.org/abs/1704.02297}{{\ttfamily arXiv:1704.02297
  [astro-ph.CO]}}.

\bibitem{Aprile:2017lqx}
{\bfseries XENON100} Collaboration, E.~Aprile {\em et~al.}, ``{Search for
  Bosonic Super-WIMP Interactions with the XENON100 Experiment},''
  \href{http://dx.doi.org/10.1103/PhysRevD.96.122002}{{\em Phys. Rev. D}
  {\bfseries 96} no.~12, (2017) 122002},
  \href{http://arxiv.org/abs/1709.02222}{{\ttfamily arXiv:1709.02222
  [astro-ph.CO]}}.

\bibitem{Aprile:2019xxb}
{\bfseries XENON} Collaboration, E.~Aprile {\em et~al.}, ``{Light Dark Matter
  Search with Ionization Signals in XENON1T},''
  \href{http://dx.doi.org/10.1103/PhysRevLett.123.251801}{{\em Phys. Rev.
  Lett.} {\bfseries 123} no.~25, (2019) 251801},
  \href{http://arxiv.org/abs/1907.11485}{{\ttfamily arXiv:1907.11485
  [hep-ex]}}.

\bibitem{Isern:1992gia}
J.~Isern, M.~Hernanz, and E.~Garcia-Berro, ``{Axion cooling of white dwarfs},''
  \href{http://dx.doi.org/10.1086/186416}{{\em Astrophys. J. Lett.} {\bfseries
  392} (1992) L23}.

\bibitem{Viaux:2013lha}
N.~Viaux, M.~Catelan, P.~B. Stetson, G.~Raffelt, J.~Redondo, A.~A.~R. Valcarce,
  and A.~Weiss, ``{Neutrino and axion bounds from the globular cluster M5 (NGC
  5904)},'' \href{http://dx.doi.org/10.1103/PhysRevLett.111.231301}{{\em Phys.
  Rev. Lett.} {\bfseries 111} (2013) 231301},
  \href{http://arxiv.org/abs/1311.1669}{{\ttfamily arXiv:1311.1669
  [astro-ph.SR]}}.

\bibitem{Skillman:2002aa}
R.~C.-P. E.~D. Skillman, ``{The ratio of blue to red supergiants in sextans a
  from hst imaging},'' \href{http://arxiv.org/abs/astro-ph/0203284}{{\ttfamily
  arXiv:astro-ph/0203284}}.

\bibitem{McQuinn:2011bb}
K.~B. McQuinn, E.~D. Skillman, J.~J. Dalcanton, A.~E. Dolphin, J.~Holtzman,
  D.~R. Weisz, and B.~F. Williams, ``{Observational Constraints on Red and Blue
  Helium Burning Sequences},''
  \href{http://dx.doi.org/10.1088/0004-637X/740/1/48}{{\em Astrophys. J.}
  {\bfseries 740} (2011) 48}, \href{http://arxiv.org/abs/1108.1405}{{\ttfamily
  arXiv:1108.1405 [astro-ph.CO]}}.

\bibitem{Friedland:2012hj}
A.~Friedland, M.~Giannotti, and M.~Wise, ``{Constraining the Axion-Photon
  Coupling with Massive Stars},''
  \href{http://dx.doi.org/10.1103/PhysRevLett.110.061101}{{\em Phys. Rev.
  Lett.} {\bfseries 110} no.~6, (2013) 061101},
  \href{http://arxiv.org/abs/1210.1271}{{\ttfamily arXiv:1210.1271 [hep-ph]}}.

\bibitem{Leinson:2014ioa}
L.~Leinson, ``{Axion mass limit from observations of the neutron star in
  Cassiopeia A},'' \href{http://dx.doi.org/10.1088/1475-7516/2014/08/031}{{\em
  JCAP} {\bfseries 08} (2014) 031},
  \href{http://arxiv.org/abs/1405.6873}{{\ttfamily arXiv:1405.6873 [hep-ph]}}.

\bibitem{Isern:2008fs}
J.~Isern, S.~Catalan, E.~Garcia-Berro, and S.~Torres, ``{Axions and the white
  dwarf luminosity function},''
  \href{http://dx.doi.org/10.1088/1742-6596/172/1/012005}{{\em J. Phys. Conf.
  Ser.} {\bfseries 172} (2009) 012005},
  \href{http://arxiv.org/abs/0812.3043}{{\ttfamily arXiv:0812.3043
  [astro-ph]}}.

\bibitem{Corsico:2012sh}
A.~Corsico, L.~Althaus, A.~Romero, A.~Mukadam, E.~Garcia-Berro, J.~Isern,
  S.~Kepler, and M.~Corti, ``{An independent limit on the axion mass from the
  variable white dwarf star R548},''
  \href{http://dx.doi.org/10.1088/1475-7516/2012/12/010}{{\em JCAP} {\bfseries
  12} (2012) 010}, \href{http://arxiv.org/abs/1211.3389}{{\ttfamily
  arXiv:1211.3389 [astro-ph.SR]}}.

\bibitem{Aver:2015iza}
E.~Aver, K.~A. Olive, and E.~D. Skillman, ``{The effects of He I $\lambda$10830
  on helium abundance determinations},''
  \href{http://dx.doi.org/10.1088/1475-7516/2015/07/011}{{\em JCAP} {\bfseries
  07} (2015) 011}, \href{http://arxiv.org/abs/1503.08146}{{\ttfamily
  arXiv:1503.08146 [astro-ph.CO]}}.

\bibitem{Hoof:2018ieb}
S.~Hoof, F.~Kahlhoefer, P.~Scott, C.~Weniger, and M.~White, ``{Axion global
  fits with Peccei-Quinn symmetry breaking before inflation using GAMBIT},''
  \href{http://dx.doi.org/10.1007/JHEP03(2019)191}{{\em JHEP} {\bfseries 03}
  (2019) 191}, \href{http://arxiv.org/abs/1810.07192}{{\ttfamily
  arXiv:1810.07192 [hep-ph]}}. [Erratum: JHEP 11, 099 (2019)].

\bibitem{Drukier:1986tm}
A.~Drukier, K.~Freese, and D.~Spergel, ``{Detecting Cold Dark Matter
  Candidates},'' \href{http://dx.doi.org/10.1103/PhysRevD.33.3495}{{\em Phys.
  Rev. D} {\bfseries 33} (1986) 3495--3508}.

\bibitem{Hogan:1988mp}
C.~Hogan and M.~Rees, ``{Axion miniclusters},''
  \href{http://dx.doi.org/10.1016/0370-2693(88)91655-3}{{\em Phys. Lett. B}
  {\bfseries 205} (1988) 228--230}.

\bibitem{Kolb:1993zz}
E.~W. Kolb and I.~I. Tkachev, ``{Axion miniclusters and Bose stars},''
  \href{http://dx.doi.org/10.1103/PhysRevLett.71.3051}{{\em Phys. Rev. Lett.}
  {\bfseries 71} (1993) 3051--3054},
  \href{http://arxiv.org/abs/hep-ph/9303313}{{\ttfamily arXiv:hep-ph/9303313}}.

\bibitem{Aprile:2015uzo}
{\bfseries XENON} Collaboration, E.~Aprile {\em et~al.}, ``{Physics reach of
  the XENON1T dark matter experiment},''
  \href{http://dx.doi.org/10.1088/1475-7516/2016/04/027}{{\em JCAP} {\bfseries
  04} (2016) 027}, \href{http://arxiv.org/abs/1512.07501}{{\ttfamily
  arXiv:1512.07501 [physics.ins-det]}}.

\bibitem{Zhang:2018xdp}
{\bfseries PandaX} Collaboration, H.~Zhang {\em et~al.}, ``{Dark matter direct
  search sensitivity of the PandaX-4T experiment},''
  \href{http://dx.doi.org/10.1007/s11433-018-9259-0}{{\em Sci. China Phys.
  Mech. Astron.} {\bfseries 62} no.~3, (2019) 31011},
  \href{http://arxiv.org/abs/1806.02229}{{\ttfamily arXiv:1806.02229
  [physics.ins-det]}}.

\bibitem{Akerib:2019fml}
{\bfseries LZ} Collaboration, D.~Akerib {\em et~al.}, ``{The LUX-ZEPLIN (LZ)
  Experiment},'' \href{http://dx.doi.org/10.1016/j.nima.2019.163047}{{\em Nucl.
  Instrum. Meth. A} {\bfseries 953} (2020) 163047},
  \href{http://arxiv.org/abs/1910.09124}{{\ttfamily arXiv:1910.09124
  [physics.ins-det]}}.

\bibitem{Aalbers:2016jon}
{\bfseries DARWIN} Collaboration, J.~Aalbers {\em et~al.}, ``{DARWIN: towards
  the ultimate dark matter detector},''
  \href{http://dx.doi.org/10.1088/1475-7516/2016/11/017}{{\em JCAP} {\bfseries
  11} (2016) 017}, \href{http://arxiv.org/abs/1606.07001}{{\ttfamily
  arXiv:1606.07001 [astro-ph.IM]}}.

\end{thebibliography}\endgroup

\end{document}